\documentclass[aps,prl,groupedaddress,]{revtex4}
\usepackage{latexsym}
\usepackage[dvips]{graphicx}

\newcommand{\eq}{\begin{equation}}
\newcommand{\feq}{\end{equation}}
\newcommand{\eqn}{\begin{eqnarray}}
\newcommand{\feqn}{\end{eqnarray}}
\newcommand{\arr}{\begin{eqnarray*}}
\newcommand{\farr}{\end{eqnarray*}}
\newcommand{\beq}{\begin{equation}}
\newcommand{\eeq}{\end{equation}}
\newcommand{\bea}{\begin{eqnarray}}
\newcommand{\eea}{\end{eqnarray}}

\def\beq{\begin{equation}}
\def\eeq{\end{equation}}
\def\feq{\end{equation}}
\def\bea{\begin{eqnarray}}
\def\eea{\end{eqnarray}}
\def\bc{\begin{displaymath}}
\def\ec{\end{displaymath}}

\begin{document}


\title{Higher Curvature Brane Corrections to the DGP Model}

\author{Mariano Cadoni}
\email{mariano.cadoni@ca.infn.it}
\author{Paolo Pani}
\email{paolo.pani@ca.infn.it}
\affiliation{Dipartimento di Fisica,
Universit\`a di Cagliari, and INFN sezione di Cagliari, Cittadella
Universitaria 09042 Monserrato, ITALY}


\begin{abstract}
We investigate the  Dvali-Gabadadze-Porrati (DGP)  model   corrected
by higher
curvature brane terms. We show that these corrections have a
dramatic impact on the spectrum of the model at the linearized level.
Owing
to the
presence of higher derivatives in the field equations very massive
ghost
excitations with  mass of order of Planck mass are generated in the
ordinary branch of the model. These excitations
describe an instability
of Minkowski vacuum with time-scale of order  of the Hubble time
$H^{-1}_{0}$.
At large distances  these tachyonic excitations are expected to
decouple
from brane-localized matter. Our modified DGP model represents
therefore a very promising framework for  solving of the
cosmological constant problem, in which Planck-scale physics is
responsible for the elementary excitations driving the accelerated
expansion of the universe, but the time-scale of the instability is
settled  by gravitational  physics at large scales.

\end{abstract}

\maketitle
Gravity
models that allow for large-distance deviations from standard
General Relativity (GR),  the so-called
Modified theories  of gravity \cite{Dvali:2000hr,ArkaniHamed:2003uy,
Dubovsky:2004sg,Kunz:2006ca}, have been widely investigated in recent
years.
The main motivation behind this interest is the hope that
the accelerated expansion of the universe
\cite{Tonry:2003zg,Spergel:2003cb,Tegmark:2003uf,Riess:2004nr} could
be explained  and the
cosmological constant problem solved  by very large distance, of
order $r\gg r_{c}=H_0^{-1}$, modifications of General Relativity
without postulating nonbaryonic forms of matter such as
dark energy. From a purely  theoretical point of view  Modified
Gravity represents a way to circumvent
Weinberg's no-go theorem on the ``old'' cosmological constant problem
\cite{Weinberg:1988cp,Carroll:2000fy} and to shift it from the realm
of short
distance particle physics to that of large distance gravitational
physics. Moreover, from the phenomenological point of view there is
plenty of room for such infrared modifications of gravitational
physics.
Our experimental knowledge
of gravity is in agreement  with GR but is limited to
distances between say $~ 10^{-3}cm - 100 MPc$.

One of the most promising  scenarios for modifications of GR in a
relativistic and general covariant framework is  the
Dvali-Gabadadze-Porrati (DGP) model \cite{Dvali:2000hr}.
The DGP model works in  the context of a
brane-world scenario (see Ref. \cite{Akama:1982jy} for a early 
proposal), in which bulk non-local effects modify the
large distance behavior of gravity  on the brane. In particular, the
DGP model allows in the so-called
\emph{self-accelerating branch} for tachyonic excitations describing
an instability of Minkowski vacuum with a time-scale of order
$\tau_{I}=r_{c}$.
Such an instability  could  represent a solution to the
cosmological constant problem. Unfortunately, it was soon realized
that the
tachyonic excitation is a  light ghost of mass $m\sim
m_{D}=r_{c}^{-1}$
gravitationally coupled to brane-localized matter
\cite{Luty:2003vm,Charmousis:2006pn}.
From  particle physics point of view the presence of such a light
ghost is simply  disastrous.

There is no compelling reason for having $\tau_{I}\sim
m_{D}^{-1}$. We would rather expect the cosmological constant problem
to be solved by a conspiracy between short-distances and
large-distances physics.
Motivated by these considerations, in this letter we explore the
possibility of a brane-world scenario in
which the the time-scale of the tachyonic instability is still of
order
$r_{c}$ ( i.e it is determined by large-distance gravitational
physics), but its mass is of order of 4D Planck-mass $M_P$.
Introducing  terms quadratic in the curvatures in
the brane part of the DGP action we show that  at the linearized
level the model
allows, in the \emph{ordinary branch}, for tachyonic
excitations
describing an instability of
Minkowski vacuum with timescale $\sim r_{c}$. Although these
excitations are ghosts, they are heavy   with mass $\sim
M_P$ and at large distances  they are expected to  decoupled
from brane-localized matter.

In the DGP model all known interactions, except gravity, are
confined on a zero-tension $(1+3)$-brane that is embedded in a
five-dimensional (5D) infinite-volume space-time (the bulk).
Although  5D brane-induced
gravity  cannot solve the old cosmological
constant problem \cite{Dvali:2000xg}, it can be used to
deepen our understanding about modified gravity
theories. The DGP action is
\beq
S=M^3\int d^5 X\sqrt{G}\;{}^{(5)}R+M_P^2\int d^4x\sqrt{-g}R
+bound.\,\,
terms\,,
\label{eq:actDGP}
\eeq
where $M$, $X$, $G$ and ${}^{(5)} R$ are bulk quantities, while
$M_P$, $x$, $g$ and $R$ are referred to the brane. In
Eq. (\ref{eq:actDGP}) boundary, Gibbons-Hawking, terms should be taken
in account to warrant the correct equations in the bulk.
Matter
fields are also localized on the brane and they are omitted here for
simplicity.\\
The relevant new  ingredient in Eq. (\ref{eq:actDGP}) is the
Einstein-Hilbert term on the brane. This 4D kinetic term, as well as
all the possible Lorentz invariant terms, can be induced via loop
corrections by brane-localized matter.\\
Using the  notation $\eta_{AB}=(+----)$, with capital indices running
on
$0,1,2,3,5$ and Greek indices running on $0,1,2,3$, the field
equations  stemming from  the action (\ref{eq:actDGP}) are,
\begin{equation}
G^{(5)}_{AB}=\frac{1}{2M^3}\delta(\text{brane})\delta^\mu_A
\delta^\nu_B(T_{\mu\nu}-2
M_P^2 G_{\mu\nu}), \label{eq:eqDGP}
\end{equation}
where $G^{(5)}_{AB}$ is the Einstein tensor built from the 5D
metric, $g^{(5)}_{AB}$, and $G_{\mu\nu}$ is the 4D Einstein tensor
built from the induced 4D metric,
$g_{\mu\nu}=\partial_\mu X^A\partial\nu X^Bg^{(5)}_{AB}$.
$X^A(x^\mu)$ defines the brane position in the bulk. We consider a
static brane, located at $y=:x^{5}=0$,  endowed with a $Z_2$-symmetry.
Eqs.  (\ref{eq:eqDGP}) are extremely difficult to solve,
only symmetric solutions are known. Deffayet
found the modified Friedman-Robertson-Walker cosmological solutions
\cite{Deffayet:2000uy}, while some approximation for the
Schwarzschild solution
has been also derived \cite{Gabadadze:2004iy,Middleton:2003yr}.

Relevant physical information about the model (spectrum of
excitations, stability) can be gained at the linearized level
by expanding the metric around the Minkowski vacuum:
$g^{(5)}_{AB}=\eta_{AB}+h_{AB}$.
The $\{\mu\nu\}$ component of Eq.
(\ref{eq:eqDGP}) in the harmonic gauge, $\partial^A
h_{AB}=\frac{1}{2}\partial_B h^A_A$, becomes
\beq
\left[M^3\partial_C\partial^C+M_P^2\delta(y)
\partial_\alpha\partial^\alpha\right]h_{\mu\nu}
=-\delta(y)(T_{\mu\nu}-\frac{1}{3}\eta_{\mu\nu}T)+
\delta(y)M_P^2\partial_\mu\partial_\nu
h_{55}\,.
\eeq
The structure of the propagator in the Fourier space is
\beq
\tilde{h}_{\mu\nu}(p,y=0)\tilde{T}^{\mu\nu}(p)=\frac{1}{M_P^2}
\frac{\tilde{T}_{\mu\nu}\tilde{T}^{\mu\nu}-\frac{1}{3}\tilde{T}^2}{p^2\pm
m_{D}p}\,,\label{eq:propDGP}
\eeq
where $p=\sqrt{p^2}$ is the square root of the \emph{Euclidean}
4-momentum, and $m_D=2M^3/M_P^2$ is the DGP crossover parameter. The
different signs in Eq. (\ref{eq:propDGP}) are related to
different boundary conditions at $y\rightarrow\infty$ and
to the  brane  embedding in the 5D bulk. They describe different
branches of solutions. The ``$+$'' sign  refers to  the
\emph{ordinary branch} and the
``$-$'' sign to the \emph{self-accelerating branch}.
The spectrum of linear excitations near the Minkowski vacuum can be
read out  from the pole structure of the propagator
(\ref{eq:propDGP}).
In the \emph{ordinary branch}  we have a massless excitation, which
does not propagate physical degrees of freedom (corresponding to a
pole with zero residue)   and a  spin 2  resonance of mass
$m=m_{D}$.  In the  \emph{self-accelerating branch} we always
have the
massless excitation and a tachyon of mass $m_{D}$.

The tensorial part of Eq.  (\ref{eq:propDGP})
 is the same as that of Pauli-Fierz (PF) theory for a massive
graviton. In fact, from a 4D point of view, gravity in the DGP model
is mediated by a
continuum of massive Kaluza-Klein modes, with no normalizable
massless graviton in the spectrum. This is a consequence of the
presence of a  flat and infinite size extra-dimension.
We have therefore a  5D (or a massive 4D) graviton
propagating 5 on-shell degrees of freedom: a helicity-2 state which is
the analog of the GR massless graviton, a helicity-1 state which does
not contribute to the propagator because of the conservation of the
stress tensor and moreover there is a helicity-0 state. The presence
of the latter state is the cause of the
van Dam-Veltman-Zakharov (vDVZ) discontinuity \cite{vanDam:1970vg}.
This is
related to the fact that in the limit $m_D\rightarrow
0$ of the  linearized approach, Newton potential is not exactly
recovered but
the Newtonian constant $G_N$ has to be renormalized.

The static potential of the DGP model  behaves therefore like
$V(r)\sim 1/r$ for $r\ll r_C=1/m_D$,
while it behaves as a truly five dimensional Newton potential,
$V(r)\sim 1/r^2$, for $r\gg r_C$. It is usually required $r_C\sim
H_0^{-1}�$ in order  to allow for IR modifications of GR.

Although the vDVZ discontinuity makes the linear approximation of the
DGP model problematic, it has been argued that nonlinearities may play
a fundamental work, restoring solutions
sufficiently close to those of GR at  solar system scales
\cite{Vainshtein:1972sx,Deffayet:2008zz,Gabadadze:2007dv,Gabadadze:2004dq}.

From a phenomenological point of view the self-accelerating branch is
the most interesting one, because it allows for cosmological de Sitter
solutions even without a 4D cosmological constant
\cite{Deffayet:2000uy}.
In fact the tachyon  corresponds to an instability of the Minkowski
vacuum
with an exponential growth time, $e^{m_D t}$. This is a very
appealing feature, since the background is modified with a curvature
term of  order $\sim m_D^2$. The new parameter, $m_D$, has to be
fine-tuned to describe the present acceleration of the
universe, thus it may seem that there is no advantage in using the DGP
model instead of  introducing a cosmological constant. But, in
the DGP case the acceleration of the universe is   a truly
dynamical effect, which is a consequence of the modified field
equations.

Unfortunately, the DGP model has been found to suffer from different
problems. There is a strong coupling  problem emerging when one tries
to
extend the results beyond the linear approximation
\cite{Rubakov:2003zb,Luty:2003vm}. At the
first non linear order the propagator gets divergent contributions.
But the most  dangerous  issue is the presence of ghosts in the
spectrum.
It has been shown  \cite{Luty:2003vm,Charmousis:2006pn} that the
tachyon in the
self-accelerating branch is a ghost of mass $m_{D}$, i.e. it has a
\emph{negative}
kinetic energy term. The
presence of
 a light gravitational ghost excitation that couples to
brane-localized
 matter  makes the 4D description meaningless and
seems to
rule out at least the
self-accelerating branch of   the DGP model as explanation for cosmic
acceleration \cite{Charmousis:2006pn}.

On the other hand the ordinary branch is ghost-free,
but in this case the Minkowski vacuum is stable and no de Sitter
solutions exist without introducing a positive cosmological constant
on the brane. This is however strictly  true only  if we consider a
brane action of the Einstein-Hilbert type. The situation could change
if we allow for other general covariant terms in the brane action, e.g
 terms depending non-linearly on  the  4D curvature tensors.
Motivated by this arguments,  we consider a DGP model with also
induced
 second order curvature terms on the brane \footnote{$R^{2}$ brane 
 corrections to the DGP model have been first investigated at the 
 qualitative level in Ref. \cite{Kiritsis:2001bc}.}:
\beq
S=M^3\int d^5 X\sqrt{G}R_{(5)}+\int d^4x\sqrt{-g}\left(M_P^2 R+\alpha
R_{\mu\nu}R^{\mu\nu}-\beta R^2\right)\,, \label{eq:actDGPmod}
\eeq
where $\alpha, \beta$ are dimensionless real parameters.

The DGP model is recovered for $\alpha=\beta=0$. It is
worth to notice that, due to the 4D Gauss-Bonnet relation
the quadratic terms in (\ref{eq:actDGPmod}) are
the most general second order terms that can be added on a
$(1+3)$-brane.
There are several  motivations for considering these corrections  to
the DGP model.
First, the same quantum corrections which induce the Einstein-Hilbert
term on the
brane, can also induce higher order terms via higher loop corrections.
Second, the introduction  of these contributions could
shed new
light on some pathological aspects of the DGP model.
Third, the action
(\ref{eq:actDGPmod}) represents the
most general
contribution at  linearized  level of higher curvature
brane terms.
This is because third (or higher) order curvature terms
do not contribute to the linearized field
equations (their contribution  to the field equations are of
${\cal{O}}(h_{\mu\nu}^{2}))$.
Last  but not least, second order curvature terms in the action
(\ref{eq:actDGPmod}) are suppressed by  $M_P^{-2}$ powers of 4D
Planck mass.
They  represent  short-distance brane corrections to the DGP
model. The action (\ref{eq:actDGPmod}) represents therefore a nice
framework to investigate simultaneously IR and UV modifications of GR.
This is the most promising setting for
solving the cosmological constant problem.

For $M=0$ the theory (\ref{eq:actDGPmod}) reduces to the higher
derivatives gravity theory first proposed by Stelle
\cite{Stelle:1977ry}.
This
model has some interesting features: it is renormalizable and it
contains massive spin 2 modes.
Nevertheless, it is well known that in
general such higher derivatives theories are pathological, due to the
unavoidable presence of ghosts. In Stelle's model the ghost has a
mass $\sim M_P/\sqrt{\alpha}$ and can be  therefore exited  at high
energy
scale. This makes the theory inconsistent at least if one requires
unitarity.

In the framework of brane-world models the unitarity requirement can
be reformulated in
a different way.
Actions like (\ref{eq:actDGP}) or (\ref{eq:actDGPmod}) have to be
considered
as low-energy effective descriptions of  some  fundamental unitary
quantum theory of  gravity, such as string theory.
At short distance, say of order of Planck length,
actions (\ref{eq:actDGP}) and (\ref{eq:actDGPmod}) are not longer
valid and the infinite series of
higher order terms should be taken into account. For instance, in the
complete fundamental theory
higher curvature terms can contribute in some form protected by
topological invariance (as in the Gauss-Bonnet case) such that
unitarity is restored.

Thus, as far as we are interested in the large distance,  IR
behavior of gravity
described by the model (\ref{eq:actDGPmod}), we
can neglect the presence of ghosts as long as they have a mass
$\sim M_P$ which is  above the UV cutoff of our effective theory and
they do not couple to brane-localized matter.\\

The modified DGP field equations derived from the action
(\ref{eq:actDGPmod}) are,
\beq
G^{(5)}_{AB}=\frac{1}{2M^3}\delta(y)\delta^\mu_A\delta^\nu_B\left(T_{\mu\nu}-2
M_P^2 G_{\mu\nu}+2 S_{\mu\nu}\right)\nonumber\,,
\eeq
with
\bea
S_{\mu\nu}&=&(\alpha-2\beta)\nabla_\mu\nabla_\nu
R-\alpha\nabla^\beta\nabla_\beta R_{\mu\nu}+\nonumber\\
&-&(\frac{\alpha}{2}-2\beta)g_{\mu\nu}\nabla^\beta\nabla_\beta
R+2\alpha R^{\alpha\beta}R_{\mu\alpha\nu\beta}+2\beta
RR_{\mu\nu}+\frac{1}{2}g_{\mu\nu}(\alpha R^{\mu\nu}R_{\mu\nu}-\beta
R^2)
\eea
Following the original computation of \cite{Dvali:2000hr}, the
linearized
equations for the $\{\mu\nu\}$ component read
\beq
\left[M^3\partial_C\partial^C+M_P^2\delta(y)
\partial_\alpha\partial^\alpha+\alpha\partial_\beta
\partial^\beta\partial_\alpha\partial^\alpha\right]
h_{\mu\nu}=-\delta(y)(T_{\mu\nu}-\frac{1}{3}\eta_{\mu\nu}T)+
\delta(y)\left[M_P^2\partial_\mu\partial_\nu+\alpha
\partial_\mu\partial_\nu\partial_\alpha\partial^\alpha\right]h_{55}.\,
\label{eq:linDGPmod}
\eeq
Notice that the linearized field equations (\ref{eq:linDGPmod}) do
not depend on  the parameter $\beta$.
This is due to the fact that in the
harmonic gauge the 4D linearized Ricci scalar is
identically zero. This means that apart from the $R$ term, only
the $(R_{\mu\nu})^2$ term  contributes to the propagator. In
particular
the introduction of $f(R)$ terms on the brane does not change the DGP
propagator (\ref{eq:propDGP}).\\
For $\alpha,\beta\neq 0$ the propagator turns out to be
\beq
\tilde{h}_{\mu\nu}(p,y=0)\tilde{T}^{\mu\nu}(p)=\frac{1}{M_P^2}
\left(\frac{\tilde{T}_{\mu\nu}\tilde{T}^{\mu\nu}-\frac{1}{3}\tilde{T}^2}
{p^2\pm
m_{D}p+\frac{\alpha}{M_P^2} p^4}\right)\,.\label{eq:propDGPmod}
\eeq
The tensorial part of the propagator is identical to that of the DGP
model and to that of  massive gravity.
But in our case the excitation
spectrum of the theory is much richer and it depends not only on the
two branches of the theory, but also on the parameter space
$(m_D,\alpha ,M_P)$. Moreover,  owing the the higher derivative terms
in Eq.
(\ref{eq:linDGPmod}) we expect the presence of ghosts in the spectrum.

As usual, the spectrum of linear excitations  can be
inferred   from the pole structure of the propagator
(\ref{eq:propDGPmod}).
In the $[\text{Re}(p),\text{Im}(p)]$ plane the propagator
(\ref{eq:propDGPmod}) has in general 4 complex poles. One is located
at $p^2=0$. As in the DGP model the residue of this pole is
zero, consistently with the absence of a normalizable 4D massless
state
in the spectrum. For $\alpha\neq0$ the other three poles can be found
solving the algebraic third order equation
\beq
p^3+\frac{M_P^2}{\alpha}p\pm\frac{m_D
M_P}{\alpha}=0\,.\label{eq:polesthird}
\eeq
Defining $\alpha_0=\frac{4}{27}\left(\frac{M_P}{m_D}\right)^2$ we
find two different regions in the parameter space: (i) for $\alpha>0$
and $\alpha<-\alpha_0$ there is one real solution and two complex
conjugate solutions, (ii) for $-\alpha_0\leq\alpha<0$ there are three
real solutions. Notice that the case $\alpha<-\alpha_0$ is
phenomenologically highly suppressed since $\alpha_0\gg 1$. In any
realistic
situation we are left with $\alpha>0$ and a small $\alpha<0$. These
two regions are discussed in detail below.

\begin{figure}[ht]
\includegraphics[width=10cm]{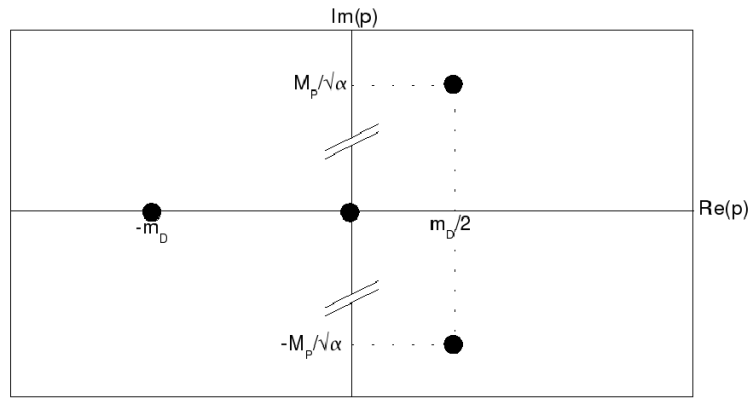}
\caption{Schematic view of the Euclidean pole structure for the DGP
model with
higher curvature brane terms in  the ordinary branch and for
$\alpha>0$. The massless particle does not propagate any degree of
freedom. This  is a consequence of the absence of normalizable
massless 4D
states in the model. The pole on the real axis corresponds to a
resonance with a typical mass $\sim m_D\sim 10^{-42}GeV$. The
resonance does not lead to any instability of the Minkowski
spacetime. The two complex poles correspond to a  massive
tachyon, $m\sim M_P/\sqrt{\alpha}\sim10^{19}GeV$.}
\label{fig:polpos}
\end{figure}

In the following we will
focus
on the ordinary branch of the theory  corresponding to the $+$ sign in
equation
(\ref{eq:propDGPmod}). Note that, due to the absence of $p^2$ terms
in Eq. (\ref{eq:polesthird}) the sum $x_1+x_2+x_3$ of the poles is
identically zero.\\
For $\alpha>0$ the solutions of Eq. (\ref{eq:polesthird}) are
\beq
x_1=-m_1\;\;\;\;\;\;\;\;\;x_{2,3}=\frac{m_1}{2}\pm m_2\,i
\,,\label{eq:solpos}
\eeq
where we have defined the ``effective'' masses
\bea
m_1=-\frac{M}{\alpha^{\frac{1}{3}}}\left[\left(-1+\sqrt{\frac{1+x}{x}}
\right)^{\frac{1}{3}}-\left(1+\sqrt{\frac{1+x}{x}}\right)^{\frac{1}{3}}
\right]\,,\label{eq:m1pos}\\
m_2=\frac{\sqrt{3}M}{2\alpha^{\frac{1}{3}}}
\left[\left(-1+\sqrt{\frac{1+x}{x}}\right)^{\frac{1}{3}}+
\left(1+\sqrt{\frac{1+x}{x}}\right)^{\frac{1}{3}}\right]\,,\label{eq:m2pos}
\eea
and $x=\frac{\alpha}{\alpha_0}$. In any phenomenologically viable
model  we have naturally $\alpha={\cal O}(1)$ and  $x$ is an extremely
small quantity ($x\sim 10^{-120}$ with
reasonable values $M_P=10^{19} GeV$, $m_D\sim H_0\sim 10^{-42} GeV$
and
$\alpha=1$). With  this
approximation the masses become
\bea
m_1&\sim& m_D\left(1-\alpha\left(\frac{m_D}{M_P}\right)^2\right)\sim
m_D\,,\label{eq:m1posapprox}\\
m_2&\sim&
\frac{M_P}{\sqrt{\alpha}}\left(1+\alpha\frac{3\sqrt{3}}{16}
\left(\frac{m_D}{M_P}\right)^2\right)\sim
\frac{M_P}{\sqrt{\alpha}}\,.\label{eq:m2posapprox}
\eea
As expected there are two independent energy scales in the model:
$m_D$,
which is the scale arising in the DGP model, and
$m_S=\frac{M_P}{\sqrt{\alpha}}$, which is exactly the mass of the
massive 4D graviton found by Stelle in the framework of gravity
theories with higher derivatives. The pole structure of the
propagator  is shown in Fig.~(\ref{fig:polpos}).
%

Interesting novel features  arises from  the investigation of
pole structure represented in Fig.~(\ref{fig:polpos}).
Similarly to  the ordinary branch of the DGP model, there is a
resonance state, corresponding to the pole $p=-m_1\sim -m_D$. This
particle couples to brane-localized matter and
does not lead to any instability of the Minkowski vacuum.
The new feature is the presence of two new
poles located at $p\sim m_D/2\pm m_S\,i$ corresponding to a
tachyon-like state with  mass $\sim m_S=M_P/\sqrt{\alpha}$. It is
easy to prove that this tachyon is a ghost. The residue of
the pole is complex and thus the norm of the state is not positive.
But this ghost is very different from that arising in the
self-accelerating branch of the
usual DGP model. In the latter case the ghost couples gravitationally
to brane-localized matter
 and has mass
$\sim m_D\sim H_0$, so it can be excited  in the IR, exactly at
the scale where we expect significant modifications from the
self-accelerating branch solution.

Conversely in the DGP model
with higher curvature brane terms  the ghost is very heavy, it has
mass  $\sim
M_P/\sqrt{\alpha}$.
Despite of its large mass, the tiny real part of tachyon pole
leads
to an instability of the Minkowski spacetime with time-scale of order
$
(m_D)^{-1}$. Self-accelerating solutions exist even in
the ordinary branch of the DGP model if one considers higher
curvature terms on the brane.
Furthermore, since in the ordinary branch the
gravitational interaction is mediated by the resonance of mass $m_D$,
we naturally expect that  at large distances the heavy particle
decouples
from brane-localized matter.

The physical explanation  of this behavior involves both 4D brane and
5D bulk physics. The tiny real (in the
Euclidean space)  contribution, of order $m_D$,   to the pole
describing the ghost has to be explained in terms of  5D non local
effects, hence as a IR effect. On the other hand the huge imaginary
contribution of order $M_P/\sqrt{\alpha}$ has to be explained has a
short distance 4D effects which modifies only the UV behavior of the
propagator.
The net result is a very
massive tachyon which has a very small and positive real part that,
as in the self-accelerating branch of the usual DGP model, leads to
an instability of the
Minkowski space. This suggests that in the DGP model with higher
curvature brane terms even the ordinary branch is self-accelerating
but without some of the pathologies arising in the self-accelerating
branch.

The essential features of the model  discussed here  seem  quite
general and in
principle are  not limited to brane-world scenarios.
They could be present, at  linearized level, in any theory of gravity 
whenever
the pole structure of the Euclidean propagator of some excitation is
determined by both IR (positive real  part) and UV (imaginary part)
effects.
The two mass regimes are so
disentangled that one can have a heavy excitation   with mass of
order
$\sqrt{\text{Re}(p)^2+\text{Im}(p)^2}\sim \text{Re}(p)$ leading to a
tiny instability of the Minkowski background.

The main difficulty  that we have to face in order to give a
consistent physical explanation of the above-described instability of
Minkowski vacuum is the ghost nature of the excitation.
The corresponding pole has a complex residue and this  leads to a
non-unitary  theory. Obviously this unitarity problem cannot be
solved in the framework of a low-energy effective theory of gravity
described by the action (\ref{eq:actDGPmod}).
One can naturally assume  the UV
completion of our theory of gravity  to be ghost free.  The ghost
could
be reabsorbed in the (infinite) series of terms of the fundamental
theory. Whether the instability due to the ghost does or does not
remain in the fundamental theory is a subtle issue that it is not
addressed here.

The presence of higher curvature term on the brane
seems to lead quite generically to an instability of the Minkowski
spacetime in the ordinary branch. There are indications that
adding a $f(R)$-term on the brane leads
to self-accelerating solutions even in the ordinary branch
\cite{mariam}. Interestingly enough, there is no trace of this
cosmological instability at the linearized level because the
propagator remains
unchanged by adding $f(R)$ brane terms. In this case instabilities of
the
Minkowski spacetime arise at the full non-linear level only.
This is another example in which nonlinear effects  in the DGP model
have
a strong impact  on  the dynamics. The
study of cosmological solutions arising from action
(\ref{eq:actDGPmod}) will be presented  elsewhere.\\
Although it is not a central issue, let us we briefly discuss the case
$-\alpha_0<\alpha<0$. Now  the solutions of Eq. (\ref{eq:polesthird})
(with the $+$ sign) are
\beq
x_1=m_0=\frac{M_P}{\sqrt{-3\alpha}}\cos\left(\frac{1}{3}
{\arctan{\sqrt{\frac{1-y}{y}}}}\right)>0,\;\;\;\;\;\;\;\;\;x_{2,3}=-
\frac{m_0}{2}\left[1\pm\sqrt{3}\tan{\left(\frac{1}{2}
{\arctan{\sqrt{\frac{1-y}{y}}}}\right)}\right]<0
\,,\label{eq:solneg}
\eeq
where $y=-\alpha/\alpha_0<1$. In this region we find two resonances
and a tachyon, all of them  with zero imaginary part. The pole
structure is
richer than in the usual  DGP case, but no new features arise in
the ordinary branch. The behavior   in  the self-accelerating branch
is qualitatively the same as that  described above. In particular,
the poles structure in the Euclidean plane is $y$-reflected. This
means that, going from one branch to the other, we just need to
change resonances into tachyons and vice versa. This general result
holds in the usual DGP case too. Always a very light tachyon-like
state appears in this branch and
it is a ghost with a mass of the order $m_D$.\\
Higher curvature brane terms do not change the large distance
behavior of the static potential of the usual DGP model. Nevertheless
they can give
short distance contributions, at energy scales of order
$M_P/\sqrt{\alpha}$.
The modified DGP model (\ref{eq:actDGPmod})   corrects  GR  both
in the IR
and in the UV. Neglecting the tensorial structure of the propagator
(\ref{eq:propDGPmod}), the static potential is
\beq
V(r)=\int dt{\int \frac{d^4p}{(2\pi)^4}
e^{ipx}\phi(p,y=0)}\,,\label{eq:potdef}
\eeq
where $\phi(p,y=0)$ is the Fourier-transformed scalar part of the
propagator (\ref{eq:propDGPmod}) on the brane. Focusing on the
phenomenologically
interesting case $\alpha>0$ the potential reads
\beq
V(r)=-\frac{1}{m_2^2\alpha\pi^2(4+9\eta^2)}\left(2\,V_1(r)-\,
\text{Re}\left[(2+3i\eta)\,V_2(r)\right]\right)\,,\label{eq:pot}
\eeq
where $\eta=m_1/m_2$ and $V_1$, $V_2$ are obtained from the standard
DGP potential
\cite{Dvali:2000hr}
\beq
V_{DGP}(r)=\frac{1}{r}\left[\sin(m_D r)\text{Ci}(m_D
r)+\frac{1}{2}\cos(m_D r)(\pi- 2\text{Si}(m_D
r))\right]\,,\label{eq:potgen}
\eeq
substituting $m_D\rightarrow m_1$ and $m_D\rightarrow m_1/2+m_2\,i$
respectively. In the large-$r$ limit ($r\gg 1/m_1,1/m_2$) the
potential (\ref{eq:pot}) becomes
\beq
V(r)\sim-\frac{1}{m_1\alpha\pi^2(4+9\eta^2)}\left[\frac{2-3\eta}{(m_2r)^2}-
\eta\pi\frac{e^{-m_2
r}}{m_2r}\left(2\cos\frac{m_1r}{2}-3\eta\sin\frac{m_1
r}{2}\right)\right]\,.\nonumber
\eeq
As expected, corrections due to higher curvature brane terms decay
exponentially.
As in the usual  DGP model the gravitational  potential is five
dimensional and its
behavior can set a limit on the DGP parameter, $m_D$. In the
intermediate region, $1/m_2\ll r\ll 1/m_1$ the potential is
\beq
V(r)\sim
-\frac{1}{m_2\alpha\pi(4+9\eta^2)}\left(\frac{1}{m_2r}-
\frac{2\eta}{\pi}(-1+\gamma-\log(m_1r))+\frac{3\eta}{\pi}\frac{1}{(m_2r)^2}+
\frac{2e^{-m_2r}}{m_2r}-\frac{3\eta^2}{2}e^{-m_2r}\right)\,,\label{eq:potINT}
\eeq
where $\gamma$ is the Euler constant. In this region one recovers
Newton potential corrected by subleading
logarithmic terms arising from the DGP bulk and by exponential and by
$r^{-2}$
corrections arising from the higher curvature brane terms. In the
small-$r$
region, $r\ll 1/m_2, 1/m_1$, the potential behaves as $\sim r$. This,
together with submillimeter table-desk experiments can put
constraints on the value
on the parameter $\alpha$.\\
In this paper we have investigated the DGP model corrected by higher
curvature brane terms. We have found that these corrections have a
dramatic effect on the linearized spectrum of the model. Ghost
excitations with  mass of order $M_P$, which describe an instability
of Minkowski vacuum with time-scales of order $H_{0}�^{-1}$are
generated.
In the IR the tachyonic excitation is expected to decouple
from brane-localized matter. Our modified DGP model represents
therefore a very promising framework for  solving the
cosmological constant problem, in which Planck scale physics is
responsible for the elementary excitations driving the accelerated
expansion of the universe, but the timescale of the instability is
settled  by the gravitational physics at large scales.

Our result gives a strong hint that short distance  corrections to
the DGP model may held the key for solving some of the open problem
of the
model. There are still some issues and consistency
checks,   which have been
not addressed in detail in this paper. The first is the particle
physics
meaning of the ghost excitation driving the instability of Minkowski
space.  The main issue here is not the ghost nature of the excitation
- it is rather natural to assume that  this problem will be solved by
the UV completion of the theory, because  any consistent quantum
theory of
 gravity, such as string theory, is expected to be unitary- but the
reliability of our low-energy approximation given by the action
(\ref{eq:actDGPmod}).  It is not clear to what extent the truncation
is consistent and if in an alternative treatment the ghost excitation
would still be present.
An other, related, issue is the reliability of perturbation theory.
Our results have been derived in the linear approximation.
It is well known that the linear approximation of the DGP model is
strongly limited because of the vDVZ discontinuity and of the strong
coupling  effect. Nonlinear effects are therefore crucial for making
the model consistent. It is not clear whether the features we have
found
at
the linearized level still persist at a full nonlinear level.
A  very important check in this context is represented by the
investigation of the cosmological solutions of the our model.
It is also important to stress that the introduction of
higher-derivatives
interactions   -also in the form
of higher curvature terms-  in the DGP model may be also very useful
for
solving the vDVZ discontinuity problem. These terms
can support a robust
implementation of the Vainshtein effect \cite{Vainshtein:1972sx} and
still allow for
self-accelerating de Sitter solutions with no light ghost
instabilities
\cite{Nicolis:2008in}.

{\bf Acknowledgements}\\
We thank G. D'Appollonio  for discussions and valuable comments.


\begin{thebibliography}{99}

\bibitem{Dvali:2000hr}
  G.~R.~Dvali, G.~Gabadadze and M.~Porrati,
  Phys.\ Lett.\  B {\bf 485} (2000) 208
  [arXiv:hep-th/0005016].

\bibitem{ArkaniHamed:2003uy}
  N.~Arkani-Hamed, H.~C.~Cheng, M.~A.~Luty and S.~Mukohyama,
gravity,''
  JHEP {\bf 0405} (2004) 074
  [arXiv:hep-th/0312099].

\bibitem{Dubovsky:2004sg}
  S.~L.~Dubovsky,
  JHEP {\bf 0410} (2004) 076
  [arXiv:hep-th/0409124].
\bibitem{Kunz:2006ca}
  M.~Kunz and D.~Sapone,
  Phys.\ Rev.\ Lett.\  {\bf 98} (2007) 121301
  [arXiv:astro-ph/0612452].

\bibitem{Tonry:2003zg}
  J.~L.~Tonry {\it et al.}  [Supernova Search Team Collaboration],
  Astrophys.\ J.\  {\bf 594} (2003) 1
  [arXiv:astro-ph/0305008].

\bibitem{Spergel:2003cb}
  D.~N.~Spergel {\it et al.}  [WMAP Collaboration],
Observations:
  Astrophys.\ J.\ Suppl.\  {\bf 148} (2003) 175
  [arXiv:astro-ph/0302209].


\bibitem{Tegmark:2003uf}
  M.~Tegmark {\it et al.}  [SDSS Collaboration],
  Astrophys.\ J.\  {\bf 606} (2004) 702
  [arXiv:astro-ph/0310725].

\bibitem{Riess:2004nr}
  A.~G.~Riess {\it et al.}  [Supernova Search Team Collaboration],
Telescope:
Evolution,''
  Astrophys.\ J.\  {\bf 607} (2004) 665
  [arXiv:astro-ph/0402512].



\bibitem{Weinberg:1988cp}
  S.~Weinberg,
  Rev.\ Mod.\ Phys.\  {\bf 61} (1989) 1.

\bibitem{Carroll:2000fy}
S.~M.~Carroll,
``The cosmological constant,''
Living Rev.\ Rel.\  {\bf 4} (2001) 1
[arXiv:astro-ph/0004075].

\bibitem{Akama:1982jy}
  K.~Akama,
  Lect.\ Notes Phys.\  {\bf 176} (1982) 267
  [arXiv:hep-th/0001113].

\bibitem{Luty:2003vm}
  M.~A.~Luty, M.~Porrati and R.~Rattazzi,
  JHEP {\bf 0309} (2003) 029
  [arXiv:hep-th/0303116].

\bibitem{Charmousis:2006pn}
  C.~Charmousis, R.~Gregory, N.~Kaloper and A.~Padilla,
  JHEP {\bf 0610} (2006) 066
  [arXiv:hep-th/0604086].

\bibitem{Dvali:2000xg}
  G.~R.~Dvali and G.~Gabadadze,
  Phys.\ Rev.\  D {\bf 63} (2001) 065007
  [arXiv:hep-th/0008054].

\bibitem{Deffayet:2000uy}
 C.~Deffayet,
Phys.\ Lett.\  B {\bf 502} (2001) 199
[arXiv:hep-th/0010186].


\bibitem{Gabadadze:2004iy}
  G.~Gabadadze and A.~Iglesias,
  Phys.\ Rev.\  D {\bf 72} (2005) 084024
  [arXiv:hep-th/0407049].

\bibitem{Middleton:2003yr}
  C.~Middleton and G.~Siopsis,
  Mod.\ Phys.\ Lett.\  A {\bf 19} (2004) 2259
  [arXiv:hep-th/0311070].

\bibitem{vanDam:1970vg}
  H.~van Dam and M.~J.~G.~Veltman,
  Nucl.\ Phys.\  B {\bf 22}, 397 (1970).

\bibitem{Vainshtein:1972sx}
  A.~I.~Vainshtein,
  Phys.\ Lett.\  B {\bf 39} (1972) 393.

\bibitem{Deffayet:2008zz}
  C.~Deffayet,
  Int.\ J.\ Mod.\ Phys.\  D {\bf 16} (2008) 2023.

\bibitem{Gabadadze:2007dv}
  G.~Gabadadze,
  Nucl.\ Phys.\ Proc.\ Suppl.\  {\bf 171} (2007) 88
  [arXiv:0705.1929 [hep-th]].

\bibitem{Gabadadze:2004dq}
  G.~Gabadadze,
  arXiv:hep-th/0408118.

\bibitem{Rubakov:2003zb}
  V.~A.~Rubakov,
  arXiv:hep-th/0303125.


\bibitem{Kiritsis:2001bc}
  E.~Kiritsis, N.~Tetradis and T.~N.~Tomaras,
  JHEP {\bf 0108} (2001) 012
  [arXiv:hep-th/0106050].

  \bibitem{Stelle:1977ry}
  K.~S.~Stelle,
  Gen.\ Rel.\ Grav.\  {\bf 9}, 353 (1978).


\bibitem{mariam} M.~Bouhmadi-L\'opez and R.~Lazkoz, work in
preparation.


\bibitem{Nicolis:2008in}
  A.~Nicolis, R.~Rattazzi and E.~Trincherini,
  arXiv:0811.2197 [hep-th].














\end{thebibliography}
\end{document}